\documentclass{article}

\usepackage{arxiv}

\usepackage[utf8]{inputenc} 
\usepackage[T1]{fontenc}    
\usepackage{hyperref}       
\usepackage{url}            
\usepackage{booktabs}       
\usepackage{amsfonts}       
\usepackage{nicefrac}       
\usepackage{microtype}      
\usepackage{lipsum}
\usepackage{graphicx}
\graphicspath{ {./images/} }

\usepackage{algorithm}
\usepackage{algorithmic}
\usepackage{amsfonts}
\usepackage{amsmath}
\newtheorem{Definition}{Definition}

\newtheorem{Property}{Property}

\newtheorem{Theorem}{Theorem}
\usepackage{multirow}

\usepackage{newfloat}
\usepackage{listings}

\usepackage{graphicx}

\usepackage{dsfont}

\usepackage{array}

\usepackage[numbers]{natbib}

\date{}

\begin{document}

\title{Deterministic-Allocation and Anonymous Joint Advertising in E-commerce Platforms}

\author{
    \textbf{Zhen Zhang}\thanks{Both authors contribute equally to this research.} \\
    \texttt{zhangzhen2023@ruc.edu.cn} \\
    Gaoling School of  \\ Artificial Intelligence, \\ Renmin University of China \\
    Beijing, China
    \And
    \textbf{Luowen Liu} \footnotemark[1] \\
    \textbf{Wanzhi Zhang} \\
    \texttt{liuluowen@meituan.com} \\
    \texttt{wanzhiz@163.com} \\
    Meituan Inc. \\
    Beijing, China
    \And
    \textbf{Zitian Guo} \\
    \texttt{ztguo@ucsd.edu}\\
    University of California, San Diego \\
    La Jolla, United States
    \And
    \\
    \textbf{Kun Huang} \\
    \texttt{huangkun13@meituan.com}\\
    Meituan Inc. \\
    Beijing, China
    \And
    \\
    \textbf{Qi Qi}\thanks{Corresponding author.} \thanks{Also with, Beijing Key Laboratory of Research on Large Models and Intelligent Governance. Also with, Engineering Research Center of Next-Generation Intelligent Search and Recommendation, MOE.} \\
    \texttt{qi.qi@ruc.edu.cn} \\ 
    Gaoling School of \\ Artificial Intelligence, \\ Renmin University of China \\
    Beijing, China
    \And
    \\
    \textbf{Qianlong Xie} \\
    \textbf{Xingxing Wang} \\
    \texttt{{xieqianlong,wangxingxing04}@}\\
    \texttt{meituan.com}\\
    Meituan Inc. \\
    Beijing, China
}

\maketitle

\begin{abstract}
With the advancement of machine learning, an increasing number of studies are employing automated mechanism design (AMD) methods for optimal auction design. However, all previous AMD architectures designed to generate optimal mechanisms that satisfy near dominant strategy incentive compatibility (DSIC) fail to achieve deterministic allocation, and some also lack anonymity, thereby impacting the efficiency and fairness of advertising allocation. This has resulted in a notable discrepancy between the previous AMD architectures for generating near-DSIC optimal mechanisms and the demands of real-world advertising scenarios. In this paper, we prove that in all online advertising scenarios, when all ad slots must be allocated, previous non-deterministic allocation AMD methods lead to the non-existence of feasible solutions in the vast majority of cases, resulting in a gap between the rounded solution and the optimal solution. Furthermore, we propose JTransNet, a transformer-based neural network architecture, designed for optimal deterministic-allocation and anonymous joint auction design. Although the deterministic allocation module in JTransNet is designed for the latest joint auction scenarios, it can be applied to other non-deterministic AMD architectures with minor modifications.
Additionally, our offline and online data experiments demonstrate that, in joint auction scenarios, JTransNet significantly outperforms the considered baselines in terms of platform revenue.
\end{abstract}


\section{Introduction}



Internet ad generates substantial revenue, amounting to tens of billions of dollars annually, for internet companies. The field of ad auction design has attracted considerable attention from both academia and industry. In particular, sponsored search auctions, a prevalent method in online ad markets, allocate ad slots and determine payments based on advertisers' bids, leading to the concept of position auctions \cite{varian2007position, varian2009online}. 

\begin{figure*}[h]
\centering
\includegraphics[width=1.04\textwidth]{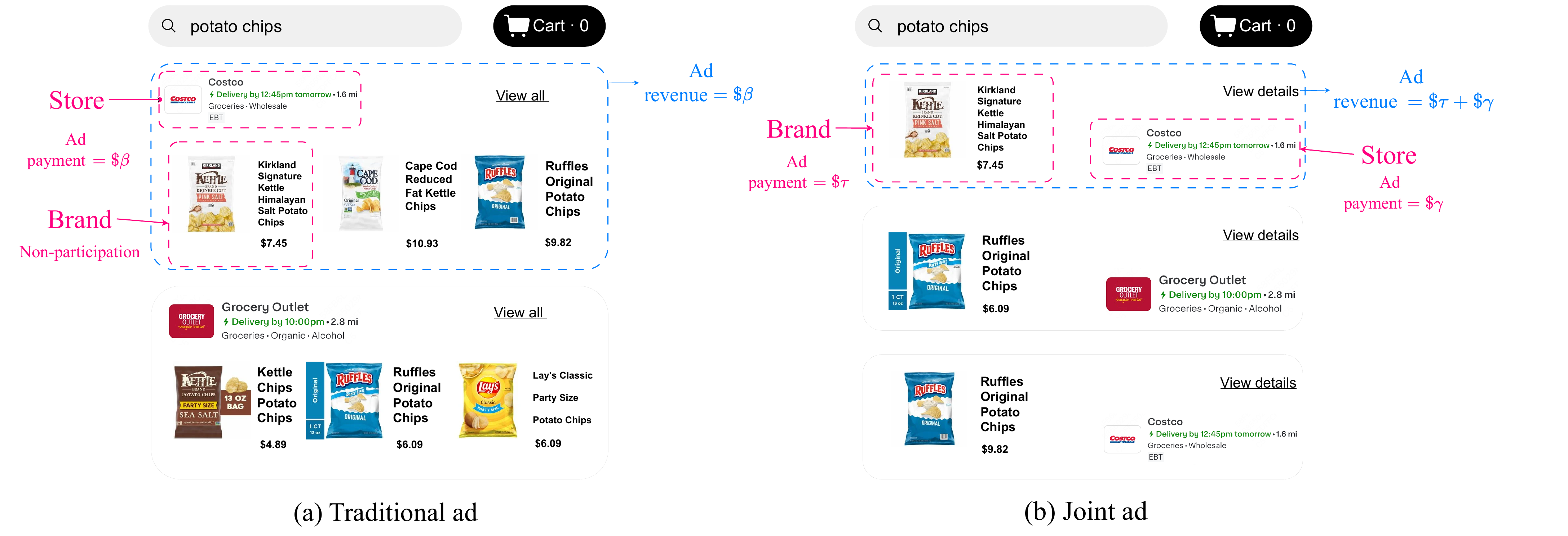}
\caption{Comparison between Traditional Ad Model and Joint Ad Model.}
\label{fig:1}
\end{figure*}

Currently, most e-commerce companies use the traditional ad model where only stores can bid for ad positions to promote their products. However, brand suppliers are equally eager to compete for these ad positions to increase the visibility of their brands. Recently, an innovative joint ad model proposed by Zhang et al. \cite{zhang2024} has effectively addressed this issue, which allows both stores and brands to bid for ad positions in ad auctions, as shown in Figure \ref{fig:1}. In the joint ad model, each ad position displays a bundle composed of a store and a brand, effectively meeting the needs of both stores and brands while also increasing the revenue of internet ad platforms, creating a win-win situation.

Nevertheless, the complexity of joint auctions, such as the difficulty in determining the payments for the store and brand within the bundle, may render most commonly used mechanisms inapplicable to joint auctions. While theoretical research has encountered bottlenecks, the paradigm of automated mechanism design (AMD) \cite{dutting2019optimal, duan2022context,zhang2024} has been proposed in recent years for optimal auction design. However, although AMD methods that satisfy near-dominant strategy incentive compatibility (DSIC) can achieve approximately optimal revenue, they have not yet been adopted in the industry, and there is no research on any such AMD architecture that has been truly deployed online. Even though these methods can achieve higher ad revenue, they do not seem to be popular in the industry.

In our attempts to deploy AMD architectures that satisfy near-DSIC online, we find that compared to widely used ad mechanisms such as GSP and VCG, all previous AMD methods generate probabilistic allocation mechanisms rather than deterministic ones where ad slots are ultimately allocated to an advertiser or a bundle with certainty. This may be the primary reason why such AMD methods have not been truly utilized in the industry. We demonstrate that in the online ad scenario, when all ad slots must be allocated, the non-deterministic probabilistic allocation results output by almost all near-DSIC AMD architectures are infeasible in the vast majority of cases, while deterministic allocation does not have this issue. Additionally, all widely adopted online ad mechanisms are deterministic, while non-deterministic probabilistic allocation mechanisms are difficult for advertisers to understand and accept, potentially leading to a significant risk of advertiser loss for internet platforms. In addition, some AMD architectures, especially those in joint scenarios \cite{zhang2024} , do not satisfy anonymity, which is a concern for advertisers. Anonymity means that the identities and the bid order of participants do not influence the auction outcomes, i.e., the auction outcomes depend only on the values of the bids and are independent of the identities and the bid order of the participants, which makes auctions fairer. Therefore, it is crucial for the true online deployment of AMD architectures in the industry to explore how to design AMD architectures that can generate optimal anonymous and deterministic allocation mechanisms that satisfy near-DSIC and individual rationality (IR).

\subsection{Main Contributions}

This paper is the first to use AMD architecture to generate optimal anonymous and deterministic-allocation mechanisms that satisfy near-DSIC and IR. We propose a novel neural network architecture, \textbf{J}oint \textbf{Trans}former-Based Neural \textbf{Net}work (JTransNet) to achieve this. JTransNet effectively tackles the unique challenges presented by anonymous and deterministic-allocation auction, which are not sufficiently addressed by existing popular neural network architectures (e.g., \cite{feng2018deep, rahme2021permutation, duan2022context, zhang2024}):

\begin{itemize}

\item \textbf{Correlated bids among the bundles.} Although JTransNet inputs the bid of each bundle to enhance the learning of bundle bid information, it ultimately converts the allocation results of each bundle for each ad slot into the allocation results for each store or brand for each ad slot, and calculates the final payment based on the independent bids of the stores and brands. This method prevents payments from depending on bundle bids, effectively mitigating the issue of correlated bids.


\item \textbf{Implementation of deterministic allocation.}
To the best of our knowledge, no previous AMD architecture designed to generate near-DSIC optimal mechanisms has accomplished deterministic allocation. This is challenging because the allocation results of deterministic mechanisms are non-differentiable, and the training of neural networks requires differentiable allocation results to compute gradients. To address this challenge, we introduce an approximately differentiable representation of the allocation results during the training phase of JTransNet to compute gradients, thereby optimizing JTransNet.

\item \textbf{Implementation of anonymity.}
In the joint auction scenario, there is currently no AMD architecture that achieves anonymity. To achieve this, we incorporate the transformer module into JTransNet and adjust its neural network structure so that changes in the order of bids do not affect the neurons corresponding to each bid, thereby generating anonymous mechanisms.

\item \textbf{Implementation of joint auction constraints.} 
In joint auctions, two constraints exist, i.e., each bundle can occupy at most one slot, and vice versa. To implement these constraints, JTransNet ensures that, in the final allocation matrix for each bundle to each ad slot, each row and each column has at most one ``1'', representing an allocation, while all other entries are ``0''.


\item \textbf{Calculation of payments.} 
Determining the individual payments of the store and brand within each bundle that obtains an ad slot is complex. To tackle this issue, JTransNet optimizes a set of parameters to scale the total expected value of  each store and brand, and the scaled values are defined as their payments. JTransNet uses a sigmoid function to calculate each parameter, guaranteeing that the payment is consistently lower than the bid, and satisfying IR.

\end{itemize}

According to our experimental results on both simulated and real data, JTransNet can achieve significantly higher revenue than the considered baseline methods with only a minimal loss in social welfare compared to the VCG mechanism that maximizes social welfare. This not only proves the effectiveness of JTransNet but also demonstrates its excellent ability to balance revenue and social welfare, making it an exceptional choice for real joint auction scenarios. Additionally, we deploy JTransNet in joint auction scenarios on a real e-commerce platform, and in online A/B tests, JTransNet substantially increases the platform's revenue. This further demonstrates the generality and scalability of JTransNet for real joint ad scenarios. As far as we know, JTransNet is the first AMD architecture for generating near-DSIC optimal mechanisms to be deployed in real industrial scenarios. Anonymity and deterministic allocation make JTransNet a truly end-to-end neural network architecture that aligns with the real ad scenarios of internet companies, making it easy to implement and deploy in the ad systems of actual internet platforms. It is worth noting that our novel model design and innovative technical methods for achieving anonymity and deterministic allocation can also be extended to other AMD architectures, which is of significant importance for the practical promotion of AMD methods.

\subsection{Related Work}

Currently, traditional ad auction mechanisms, such as the Generalized Second Price (GSP) auction \cite{varian2007position, edelman2007internet}, have been widely adopted in the industry. The simplicity and comprehensibility of GSP are the main reasons for its prevalence in practical ad markets \cite{lucier2012revenue, caragiannis2011efficiency, gomes2009bayes}. Subsequently, more and more variants of GSP have been proposed, such as the introduction of the ``squashing'' concept \cite{lahaie2007revenue}, and the simultaneous introduction of ``reserve price'' and ``squashing'' concepts \cite{thompson2013revenue}.
However, in joint auctions, it is difficult to define the equilibrium of GSP. Another important mechanism is the theoretically optimal Myerson mechanism \cite{myerson1981optimal}, which is used for single, indivisible item auctions. However, the correlated bids in joint auctions render the Myerson mechanism invalid. Additionally, the Vickrey-Clarke-Groves (VCG) mechanism \cite{vickrey1961counterspeculation, clarke1971multipart, groves1973incentives} aims to design mechanisms that optimize social welfare while satisfying DSIC and IR. There are also many other variants of the VCG mechanism, such as mixed bundling auctions \cite{jehiel2007mixed} and mixed-bundling auctions that include reserve prices \cite{tang2012mixed}. As far as we know, VCG is the only traditional mechanism that is applicable to joint auctions, but it does not perform well in terms of revenue. In this paper, we propose JTransNet to generate an optimal deterministic-allocation mechanism that satisfies near-DSIC and IR.


In contrast to the bottlenecks in theoretical research, the paradigm of automated mechanism design (AMD) \cite{conitzer2002complexity, conitzer2004self, sandholm2015automated} has been proposed for multi-item auction design with the development of machine learning. 
Currently, there are three mainstream approaches in the field of AMD: menu-based AMD methods, affine maximizer auctions (AMA) methods \cite{curry2022differentiable, duan2023scalable}, and RegretNet-based \cite{dutting2019optimal} modeling methods. Menu-based AMD methods can achieve deterministic allocation and exact incentive compatibility (IC), but these methods either cannot handle multi-agent multi-item auctions, such as MenuNet \cite{shen2018automated} and RochetNet \cite{dutting2019optimal} or cannot achieve higher revenue than near-IC methods, such as GemNet \cite{wang2024gemnet}. AMA \cite{curry2022differentiable, duan2023scalable} is an improved version of VCG that achieves exact-IC; however, the class of mechanisms it can represent is limited. RegretNet \cite{dutting2019optimal} is a foundational work in the field of AMD, which is the first neural network architecture for near-DSIC optimal multi-item auctions.
Subsequently, many extension works based on RegretNet have been proposed, such as budget constraints \cite{feng2018deep} and human preferences \cite{peri2021preferencenet}.
The permutation-equivariant architecture EquivariantNet \cite{rahme2021permutation} is proposed for symmetric auction design. Duan et al. \cite{duan2022context} design the CITransNet architecture based on Transformer for contextual auction scenarios. Ivanov et al. \cite{ivanov2022optimal} propose the RegretFormer architecture and improve the loss function. 
Zhang et al. \cite{zhang2024} proposes the first AMD architecture suitable for joint auctions, JRegNet, for optimal joint auction design. 
However, whether it is JRegNet or any previous AMD architecture based on RegretNet modeling that aims to generate near-DSIC optimal mechanisms, the encoded auction mechanisms are not deterministic allocations but non-deterministic probabilistic allocations. This brings significant challenges to the practical implementation and deployment of all AMD architectures that aim to generate near-DSIC optimal mechanisms in real-world industries. Additionally, JRegNet is not anonymous, meaning that simply changing the order of bids results in different allocation and payment outcomes under the JRegNet-encoded mechanism.  
This makes the auction very unfair and unconvincing. In addition, Aggarwal et al. \cite{aggarwal2024selling} investigate the joint advertising scenario from a theoretical perspective, modeling it as a decision-making problem over multiple periods. Building upon the joint advertising model, Zhang et al. \cite{zhang2025hybrid} further propose the hybrid advertising model and design the HRegNet architecture for optimal hybrid auction design.
Ma et al. \cite{ma2024joint} propose JAMA, which is based on AMA and suitable for joint advertising models. However, the revenue of JAMA is not high. Our proposed JTransNet solves all these problems and achieves near-optimal revenue in joint auctions.

\section{Preliminaries}
\label{sec:pre}

In this section, we formulate the joint auction scenario and convert the optimal auction design problem in this scenario into a learning problem.

\subsection{Joint ad model}

In the joint ad model, each user search corresponds to an ad auction, after which the platform presents an interface with $K$ ad slots. Each ad slot $k \in \mathbb{K}=\{1, \ldots, K\}$ has a Click-Through Rate (CTR) denoted by $\alpha_{k}$, with the assumption that, w.l.o.g., $1 > \alpha_1 \geq \cdots \geq \alpha_K > 0$.

In the joint auction phase, $m$ brands and $n$ stores participate, with each ad slot allocated to a bundle consisting of one brand and one store. Let $M = \{1, \ldots, m\}$ represent the set of brands, and $N = \{1, \ldots, n\}$ represent the set of stores. The joint relationship matrix $\{\mathbf{1}_{ij}\}_{i \in M, j \in N}$ indicates whether a specific brand and store can form a bundle. If $\mathbf{1}_{ij} = 1$, it signifies that brand $i$ and store $j$ have a selling relationship, allowing them to be bundled together. We use $C$ to represent the total number of bundles. Since, in general, the number of ads recalled by the ad platform is much greater than the number of ad slots, we assume that $C$ is greater than $K$. Moreover, in the ad scenario, two feasibility constraints must be satisfied: (a) a bundle can be assigned to no more than one slot; and (b) a single slot can be assigned to no more than one bundle.




We use $v_{i \cdot}$ (or $v_{\cdot j}$) to represent the value per click for brand $i$ (or store $j$). Each advertiser keeps their value information private. Assume that $v_{i \cdot}$ (or $v_{\cdot j}$) is drawn from a known distribution and belongs to the set $V_{i \cdot}$ (or $V_{\cdot j}$). We denote the value profile as $\boldsymbol{v} = (v_{1 \cdot}, \ldots, v_{m \cdot}, v_{\cdot 1}, \ldots, v_{\cdot n}) \in \mathbb{V}$, where $\mathbb{V} = V_{1 \cdot} \times \cdots \times V_{m \cdot} \times V_{\cdot 1} \times \cdots \times V_{\cdot n}$. The value profile excluding brand $i$ is denoted by $\boldsymbol{v}_{-i \cdot} = (v_{1 \cdot}, \ldots, v_{(i-1) \cdot}, v_{(i+1) \cdot}, \ldots, v_{\cdot n})$ (a similar definition applies for each store $j$). For each brand $i$ (or store $j$), the bid price is denoted as $b_{i \cdot}$ (or $b_{\cdot j}$). Similarly, the corresponding bid profile is represented by $\boldsymbol{b}$ and $\boldsymbol{b}_{-i \cdot}$ (or $\boldsymbol{b}_{\cdot -j}$).

We represent the joint auction mechanism as $\mathcal{M}(g,p)$, which comprises two components: the allocation rule $g=\big((g_{i \cdot})_{i \in M},$ $ (g_{\cdot j})_{j \in N} \big)$ and the payment rule $p=\big((p_{i \cdot})_{i \in M}, (p_{\cdot j})_{j \in N} \big)$. We use $C_{i \cdot}$ to represent the sets of all bundles containing brand $i$, and \( i(c) \) is used to denote the indices of the brand included in bundle \( c \).
The term $g_{i \cdot}(\boldsymbol{b}) = \sum_{c \in \mathbb{C}_{i \cdot}} \sum_{k=1}^{K}{a_{i(c) \cdot k}(\boldsymbol{b}) \alpha_k}$ represents the expected CTR for brand $i$, where \( a_{i(c) \cdot k} \) denotes the allocation to position \( k \) that brand \( i(c) \) can obtain in bundle \( c \). The calculation of \( g_{\cdot j}(\boldsymbol{b}) \) is similar.
The payment rules \((p_{i \cdot})_{i \in M} : \mathbb{V} \to \mathbb{R}^{+} \cup \{0\}\) and \((p_{\cdot j})_{j \in N} : \mathbb{V} \to \mathbb{R}^{+} \cup \{0\}\) denote the expected payment by brand \( i \) and store \( j \), respectively.
In the joint ad model, a bundle can occupy no more than one slot, and a slot can be assigned to no more than one bundle.


The utility for each brand $i$ is defined as $u_{i \cdot}(v_{i \cdot} ; \boldsymbol{b}) = v_{i \cdot}(g_{i \cdot}(\boldsymbol{b})) - p_{i \cdot}(\boldsymbol{b})$(a similar definition applies for each store $j$). In this paper, our goal is to design anonymous and deterministic-allocation mechanisms that satisfy DSIC and IR. 
The properties of deterministic allocation and  anonymity are outlined in Properties 1 and 2,  respectively, while DSIC and IR are defined in Definitions 1 and 2. According to Theorem 1, compared to non-deterministic allocation, deterministic allocation is of significant importance for the online deployment of AMD architectures in real advertising scenarios, and the proof of Theorem 1 is provided in Appendix \ref{ap1}.


\begin{Property}[Deterministic allocation]
For a joint auction, deterministic allocation is satisfied if each ad slot is deterministically allocated with a probability of $1$ to a bundle composed of a brand and a store.
\end{Property}

\begin{Property}[Anonymity]
A joint auction satisfies anonymity if and only if 
the joint auction outcomes depend only on the values of the bids and are independent of the identities and the bid order of the participants.
\end{Property}


\begin{Theorem}
The non-deterministic allocation output by all mainstream AMD architecture may not be feasible if all ad slots must be allocated.
\end{Theorem}


\begin{Definition}[DSIC]
A joint auction satisfies dominant strategy incentive compatibility if, for every brand and store, reporting their true valuation maximizes their utility, regardless of others' reports, i.e.,
$$u_{i \cdot}(v_{i \cdot}; (v_{i \cdot}, \boldsymbol{b}_{-i \cdot})) \ge u_{i \cdot}(v_{i \cdot}; (b_{i \cdot}, \boldsymbol{b}_{-i \cdot})), $$
$$\forall i \in M, \forall v_{i \cdot} \in V_{i \cdot}, \forall b_{i \cdot} \in V_{i \cdot}, \forall \boldsymbol{b}_{-i \cdot} \in \mathbb{V}_{-i \cdot},$$
and
$$u_{\cdot j}(v_{\cdot j}; (v_{\cdot j}, \boldsymbol{b}_{-\cdot j})) \ge u_{\cdot j}(v_{\cdot j}; (b_{\cdot j}, \boldsymbol{b}_{-\cdot j})), $$
$$\forall j \in N, \forall v_{\cdot j} \in V_{\cdot j}, \forall b_{\cdot j} \in V_{\cdot j}, \forall \boldsymbol{b}_{-\cdot j} \in \mathbb{V}_{-\cdot j}.$$
\end{Definition}

\begin{Definition}[IR]
A joint auction satisfies individual rationality if every brand and store can ensure a non-negative utility, as long as they truthfully report their values, i.e.,
$$u_{i \cdot}(v_{i \cdot}; (v_{i \cdot}, \boldsymbol{b}_{-i \cdot})) \geq 0, \quad \forall i \in M, \forall v_{i \cdot} \in V_{i \cdot}, \forall \boldsymbol{b}_{-i \cdot} \in \mathbb{V}_{-i \cdot},$$
and 
$$u_{\cdot j}(v_{\cdot j}; (v_{\cdot j}, \boldsymbol{b}_{-\cdot j})) \geq 0, \quad \forall j \in N, \forall v_{\cdot j} \in V_{\cdot j}, \forall \boldsymbol{b}_{-\cdot j} \in \mathbb{V}_{-\cdot j}.$$
\end{Definition}

In addition, we define the platform's expected revenue as: 
$$ rev := \mathbb{E}_{\boldsymbol{v} \sim F}\big[\sum_{i=1}^{m}{p_{i \cdot}(\boldsymbol{v})} + \sum_{j=1}^{n}{p_{\cdot j} (\boldsymbol{v})} \big],$$
i.e., the expected sum of all bidders' payments, where $F$ represents the joint value distribution of all bidders containing brands and stores. Our objective is to design an anonymous and deterministic-allocation mechanism satisfying the DSIC and IR while maximizing the platform's expected revenue, i.e., optimal anonymous and deterministic-allocation joint auction design.

\subsection{Optimal Auction Design to a Learning Problem}

We transform the optimal anonymous and deterministic-allocation joint auction design into a learning problem. First, we define ex-post regret to implement the DSIC constraint. The ex-post regret for brand $i$ is defined as follows (a similar definition applies for each store $j$):
$$
rgt_{i \cdot}(\boldsymbol{v})=\mathbb{E}_{v \sim F}[\max_{v_{i \cdot}' \in V_{i \cdot}}{u_{i \cdot} (v_{i \cdot}; (v_{i \cdot}', \boldsymbol{v}_{-i \cdot}))} - u_{i \cdot} (v_{i \cdot};\boldsymbol{v})] \text{\hfill .}
$$
$rgt_{i \cdot}(\boldsymbol{v})$ represents the maximum utility increase that brand $i$ can achieve by misreporting, while keeping others' bids fixed. Therefore, the DSIC constraint is satisfied if and only if $rgt_{i \cdot}(\boldsymbol{v}) = 0$ for each brand $i$ and $rgt_{\cdot j}(\boldsymbol{v}) = 0$ for each store $j$. Building on the definition of ex-post regret, the optimal auction design problem can be formulated as:
\begin{align}
\label{jj2}
\min_{(g,p) \in \mathcal{M}} \quad & -\mathbb{E}_{v \sim F}[\sum_{i=1}^{m}{p_{i \cdot}(\boldsymbol{v})} + \sum_{j=1}^{n}{p_{\cdot j}(\boldsymbol{v})}] \\
\text {s.t.} \quad & rgt_{i \cdot}(\boldsymbol{v}) = 0, \quad i = 1, \ldots, m,  \notag \\
& rgt_{\cdot j}(\boldsymbol{v}) = 0, \quad j = 1, \ldots, n  \notag
\end{align}
where $\mathcal{M}$ represents the set of anonymous and deterministic-allocation mechanisms that satisfy IR. Solving this optimization problem is generally difficult \cite{conitzer2002complexity, conitzer2004self}. To address this challenge, we introduce a parameter $w \in \mathbb{R}^{d}$ to parameterize the auction mechanism as $\mathcal{M}^w (g^w, p^w) \subseteq \mathcal{M}(g, p)$. As a result, our focus shifts to computing the mechanism $\mathcal{M}^w (g^w, p^w)$ that maximizes the expected revenue, given by $\mathbb{E}_{v \sim F}[\sum_{i=1}^{m}{p_{i \cdot}^w(\boldsymbol{v})} + \sum_{j=1}^{n}{p_{\cdot j}^w (\boldsymbol{v})}]$, while ensuring it satisfies the DSIC and IR conditions by adjusting the parameters $w$.

Given a sample $\mathcal{L}$ containing $L$ value profiles drawn from the joint distribution $F$, we estimate the empirical ex-post regret for brand $i$ (similarly for store $j$) under the mechanism $\mathcal{M}^w (g^w, p^w)$ by:
\begin{align}
\label{j2}
\widehat{r g t}_{i \cdot}(w)  = & \frac{1}{L} \sum_{l=1}^{L}[\max_{v_{i \cdot}^{\prime} \in V_{i \cdot}} u_{i \cdot}^w (v_{i \cdot}^{(l)}; (v_{i \cdot}^{\prime}, \boldsymbol{v}_{-i \cdot}^{(l)}))  \notag\\ 
& - u_{i \cdot}^w (v_{i \cdot}^{(l)}; \boldsymbol{v}^{(l)})] \text{\hfill .}
\end{align}
Using the sample $\mathcal{L}$, optimization (\ref{jj2}) can be expressed as:
\begin{align}
\label{j1}
\min_{w \in \mathbb{R}^{d}} \quad & -\frac{1}{L} \sum_{l=1}^{L}{[\sum_{i=1}^{m}{p_{i \cdot}^w(\boldsymbol{v}^{(l)})} + \sum_{j=1}^{n}{p_{\cdot j}^w(\boldsymbol{v}^{(l)})}]} \\
\text {s.t.} \quad & \widehat{rgt}_{i \cdot}(w) = 0, \quad i = 1, \ldots, m, \notag\\
& \widehat{rgt}_{\cdot j}(w) = 0, \quad j = 1, \ldots, n \text{\hfill .} \notag
\end{align}
Additionally, we guarantee the IR condition via the network architecture, which will be described in the next section.


\section{JTransNet}
\label{sec:JA}

After transforming the optimal joint auction design into a learning problem, in this section, we propose an end-to-end neural network, \textbf{J}oint \textbf{Trans}former-Based Neural \textbf{Net}work (JTransNet) to generate the optimal anonymous and deterministic-allocation joint auction mechanism, which satisfies near-DSIC and IR.













\begin{figure*}[ht!]
\centering
\includegraphics[width=1.12\textwidth]{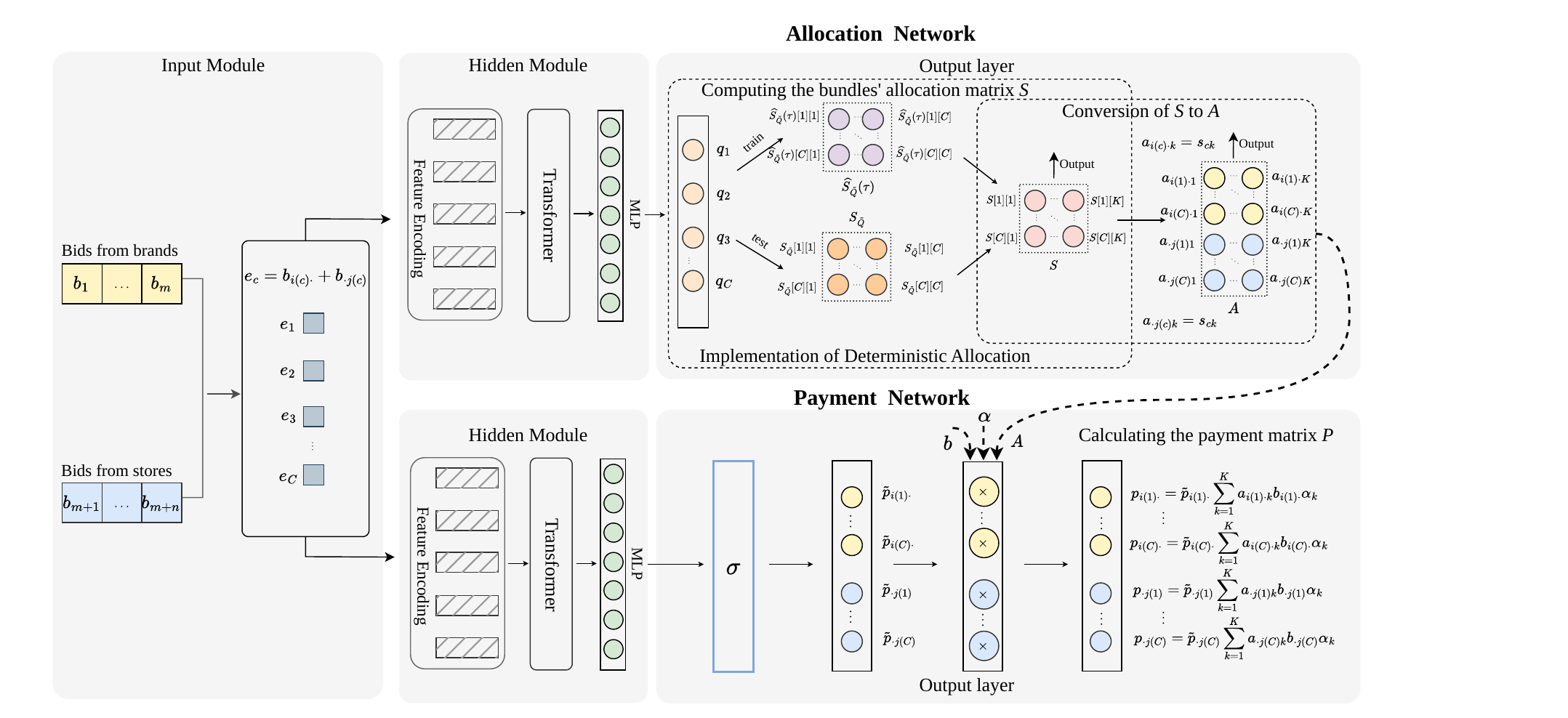}
\caption{The JTransNet architecture for the setting of $m$ brands , $n$ stores, $K$ ad slots, and $C$ bundles composed of brands and stores.}
\label{fig:3}
\end{figure*}

\subsection{The Architectures of JTransNet}
\label{sec:JA111}


In this subsection, we present the structure of JTransNet, as depicted in Figure \ref{fig:3}, and provide a detailed explanation of the key modules that constitute JTransNet.






\textbf{Input of JTransNet.} JTransNet takes $b_{i \cdot}$, $b_{j \cdot}$, $\alpha_{k}$, and $e_{c}$ as inputs, defined as:


\begin{equation}
e_c = b_{i(c) \cdot} + b_{\cdot j(c)}, \quad \forall c \in \mathbb{C} = \{1, \ldots, C\},
\label{me1}
\end{equation}
where \( i(c) \) and \( j(c) \) denote the indices of the brand and the store, respectively, included in bundle \( c \), and \( e_c \) is the bid for bundle \( c \).




\textbf{Implementation of anonymity.} We incorporate the transformer structure without positional encoding into the hidden modules and make corresponding adjustments to JTransNet to ensure that JTransNet satisfies anonymity, as shown in Figure \ref{fig:3}.




\textbf{Computing the bundles' allocation matrix $S$.} There are several constraints that need to be implemented through the bundles’ allocation matrix $S$.
Therefore, JTransNet begins by computing the bundles' allocation matrix $S$, followed by determining the bidders' allocation matrix $A$ based on $S$. 
The matrix $S$ contains the allocation probabilities for each bundle across different ad positions. Specifically, the allocation of bundle $c$ to position $k$ is denoted as $s_{c k}$ within matrix $S$. 
In our scenario, two primary constraints must be satisfied by matrix $S$: (a) a bundle can occupy no more than one slot, expressed as $\sum_{k=1}^{K} {s}_{c k} \leq 1$ for all $c \in \mathbb{C}$; and (b) a single slot can be assigned to no more than one bundle, expressed as $\sum_{c=1}^{C} {s}_{c k} \leq 1$ for all $k \in \mathbb{K}$.
In JTransNet, as illustrated by the light yellow rectangle in Figure \ref{fig:3}, the input to this part is the vector $Q=(q_1,q_2, \ldots, q_C)$, which is derived from forward propagation. The vector $Q$ contains the scores for each bundle composed of brands and stores. 
Because JTransNet aims to achieve deterministic allocation rather than non-deterministic probabilistic allocation, we achieve all the constraints by constructing the matrix \( S \) as a \(0\text{-}1\) matrix where the sum of the elements in each row and each column is less than or equal to $1$. 
To design $S$ as the desired matrix, we need to perform some transformations on $Q$ to obtain $S$. First, we calculate the index vector \(\tilde{Q}\) by sorting the elements of \( Q \) in descending order. For example, if  $Q = (9, 2, 10)$, then $\tilde{Q} = (3, 1, 2)$, where $\tilde{Q}_1 = 3$ indicates that the index of the largest number in $Q$ is 3; $\tilde{Q}_2 = 1$ indicates that the index of the second largest number is 1; and $\tilde{Q}_3 = 2$ indicates that the index of the third largest number is 2. Next, we calculate $S_{\tilde{Q}} \in\{0,1\}^{C \times C}$ through the formula as follows:

\begin{align*}
S_{\tilde{Q}}[k][c]=\begin{cases}
1 & \text{if } c=\tilde{Q}[k] \\
0 & \text{otherwise}
\end{cases} \quad \text{, for } \forall k \in \mathbb{C}, \, \forall c \in \mathbb{C} ,
\end{align*}
where $S_{\tilde{Q}}[k][c]=1$ indicates that the index corresponding to the $k$-th largest score in $Q$ is $c$. Furthermore, an alternative formula for calculating \( S_{\tilde{Q}}[k][c] \) is \cite{grover2019stochastic}:

\begin{align*}
S_{\tilde{Q}}[k][c] = \begin{cases}
1 & \text{if } c = \operatorname{argmax} \left[(C+1-2k) \mathbf{Q} - R_{\mathbf{Q}} \mathds{1}\right] \\
0 & \text{otherwise},
\end{cases}
\end{align*}
$\text{for all } k \in \mathbb{C} \text{ and } c \in \mathbb{C}$, where $R_{\mathbf{Q}}[k][c]=\left|Q_{k}-Q_{c}\right|$ and $\mathds{1}$ denotes all bundles' column vector. However, because the argmax function is not differentiable, we can use the softmax operator to replace the argmax \cite{grover2019stochastic}, resulting in a continuous \(\widehat{S}_{\tilde{Q}}(\tau) \in \mathbb{R}^{C \times C} \):

\begin{align*}
\widehat{S}_{\tilde{Q}}[k][:](\tau)=\operatorname{softmax} \left[\left((C+1-2 k) \mathbf{Q}-R_{\mathbf{Q}} \mathds{1}\right) / \tau\right] ,
\end{align*}
for all $k \in \mathbb{C}$, where \( \tau > 0 \)  determines the precision of the approximation, and as \( \tau \) approaches 0, \( \widehat{S}_{\tilde{Q}} \) converges to \( S_{\tilde{Q}} \).
Since neural network training requires $S$ to be differentiable, \(\widehat{S}_{\tilde{Q}}(\tau) \in \mathbb{R}^{C \times C} \) is used during training to calculate $S \in [0,1]^{C \times K}$:


\begin{align*}
S[c][k]=\widehat{S}_{\tilde{Q}}[k][c] \, \, \, \, \forall k \in \mathbb{K}, \, \forall c \in \mathbb{C} .
\end{align*}
During testing, we use the exact \( S_{\tilde{Q}} \in \{0,1\}^{C \times C} \) to compute \( S \):

\begin{align*}
S[c][k]=\begin{cases}
1 & \text{if } S_{\tilde{Q}}[k][c]= 1 \\
0 & \text{otherwise}
\end{cases} \, , \forall k \in \mathbb{K}, \, \forall c \in \mathbb{C} .
\end{align*}
When testing, \( S[c][k] = 1 \) means that the \( c \)-th bundle obtains the \( k \)-th ad slot; otherwise, it does not.

\textbf{Conversion of allocation matrix $S$ to $A$.} The bidders' allocation matrix \( A \) contains the allocation of all brands and stores for each ad position. 
In matrix \( A \), the allocation to position \( k \) that brand \( i(c) \) (or store \( j(c) \)) can obtain in bundle \( c \) is denoted as \( a_{i(c) \cdot k} \) (or \( a_{\cdot j(c) k} \)).  
These values, $a_{i(c) \cdot k}$ and $a_{\cdot j(c) k}$, are computed based on the bundles' allocation matrix $S$ through:

\begin{align*}
\left\{\begin{array}{l}
a_{i(c) \cdot k} = s_{c k},\quad \forall c \in \mathbb{C}, \, \forall k \in \mathbb{K}, \\
a_{\cdot j(c) k} = s_{c k}, \quad \forall c \in \mathbb{C}, \, \forall k \in \mathbb{K}.
\end{array}\right.
\end{align*}





\textbf{Calculating the payment matrix $P$.} As illustrated in Figure \ref{fig:3}, after obtaining the bidders' allocation matrix $A$, the next step is to compute the payment matrix $P$. To do this, matrix $A$ is passed through the payment network. 
We use \( p_{i(c)\cdot} \) and \( p_{\cdot j(c)} \) in \( P \) to denote the payment of brand \( i(c) \) and store \( j(c) \) in bundle \( c \), respectively, calculated as:  
 
\begin{align*}
\left\{\begin{array}{l}
p_{i(c) \cdot}=\tilde{p}_{i(c) \cdot} \sum_{k=1}^{K} a_{i(c) \cdot k} b_{i(c) \cdot} \alpha_k,\quad \forall c \in \mathbb{C}, \\
p_{\cdot j(c)}=\tilde{p}_{\cdot j(c)} \sum_{k=1}^{K} a_{\cdot j(c) k} b_{\cdot j(c)} \alpha_k, \quad \forall c \in \mathbb{C},
\end{array}\right.
\end{align*}
where \(\tilde{p}_{i(c) \cdot} \in[0,1]\) and \(\tilde{p}_{\cdot j(c)} \in[0,1]\) are computed using the sigmoid function, as illustrated by the blue rectangle in Figure \ref{fig:3}. $C_{i \cdot}$ and $C_{\cdot j}$ represent the sets of all bundles containing brand $i$ and store $j$, respectively. The total payments of brand \( i \) and store \( j \) can be expressed as \( p_{i \cdot} = \sum_{c \in \mathbb{C}_{i \cdot}} p_{i(c) \cdot} \) and \( p_{\cdot j} = \sum_{c \in \mathbb{C}_{\cdot j}} p_{\cdot j(c)} \), respectively.  
When DSIC is ensured, since the utility of the brand $i$ is $u_{i \cdot}= \sum_{c \in \mathbb{C}_{i \cdot}} (\sum_{k=1}^{K} a_{i(c) \cdot k} b_{i(c) \cdot} \alpha_k - p_{i(c) \cdot})$, and because $\tilde{p}_{i(c) \cdot} \in [0,1]$, each brand's utility must be non-negative, fulfilling IR. The process of proving IR for stores is similar. 

\subsection{Training of JTransNet}
\label{ap333}


In this subsection, we describe the training processes of JTransNet.



\textbf{Transforming constrained optimization into unconstrained optimization.} We first require an unconstrained optimization objective as the loss function to train JTransNet. The augmented Lagrangian method is applied to transform the constrained optimization problem (\ref{j1}) into an unconstrained one, as follows:

\begin{align*}
\mathcal{C}_{\rho}(w ; \boldsymbol{\lambda})= &-\frac{1}{L} \sum_{\ell=1}^{L}\left[\sum_{i=1}^{m} p_{i \cdot}^{w}\left(\boldsymbol{v}^{(\ell)}\right)+\sum_{j=1}^{n} p_{\cdot j}^{w}\left(\boldsymbol{v}^{(\ell)}\right)\right]  \\
& + \sum_{i=1}^{m}{\lambda_{i \cdot} \widehat{rgt}_{i \cdot}(w)} + \sum_{j=1}^{n}{\lambda_{\cdot j} \widehat{rgt}_{\cdot j}(w)} \\
& + \frac{\rho}{2}\sum_{i=1}^{m}{(\widehat{rgt}_{i \cdot}(w))^2} +\frac{\rho}{2} \sum_{j=1}^{n}{(\widehat{rgt}_{\cdot j}(w))^{2}} \text{\hfill ,}
\end{align*}
where $\rho>0$ denotes the penalty factor, and  $\boldsymbol{\lambda} \in \mathbb{R}^{m+n}$ represents the Lagrange multiplier. 

After obtaining the loss function, we can perform the training of JTransNet. We use \( t \in \{1, \cdots, T\} \) to denote the training iteration, where \( T \) is the total number of training iterations. In each training iteration \( t \), a mini-batch \( \mathcal{Y}_t = \{v^{(1)}, \cdots, v^{(Y)}\} \) of size \( Y \) is selected from the training set for training.

In the course of training, the optimal misreports also need to be calculated so that regret can be maximized during training.

\textbf{Computing the optimal misreports.} The regret of JTransNet is calculated by enumeration \cite{liao2022nma}. In this enumeration method, we calculate the misreport using a coefficient $\mu \in \Gamma$, where $\Gamma$ is a subset of $\mathbb{R}_{\geq 0}$, and the misreport of brand $i$  is given by $\min(\mu v_{i\cdot}, v_{i\cdot}^{\max})$ (similar calculation process for store $j$), where $v_{i\cdot}^{\max}$ denotes the maximum value in set $V_{i\cdot}$. 
For each brand and store, within the coefficient set $\Gamma$, the coefficient $\mu \in \Gamma$ that maximizes the utility of that brand or store is used to compute the misreport and regret for that brand or store, where the misreport for brand $i$ is calculated as $\min(\mu v_{i\cdot}, v_{i\cdot}^{\max})$, and analogously for store $j$.

\textbf{Updating the Lagrange multipliers.} In addition, during the training process, after every fixed number of iterations, the Lagrange multipliers are updated according to the following formula:
\begin{align*}
\left\{\begin{array}{l}
\lambda_{i \cdot }^{t+1}=\lambda_{i \cdot }^{ {t }}+\rho_{t} \widetilde{r g t}_{i \cdot }\left(w^{ {t+1 }}\right) , \quad \forall i \in M\text{\hfill ,}\\
\lambda_{\cdot j }^{ {t+1 }}=\lambda_{\cdot j }^{ {t }}+\rho_{t} \widetilde{r g t}_{\cdot j}\left(w^{ {t+1 }}\right) , \quad \forall j \in N \text{\hfill ,}
\end{array}\right.
\end{align*}
where $\widetilde{r g t}_{i \cdot}(w)$ and $\widetilde{r g t}_{\cdot j}(w)$ indicate the empirical regret of minibatch \(\mathcal{Y}_t\) computed using Formula (\ref{j2}).

\textbf{Backpropagation.} 
Afterwards, the loss function \({C}_{\rho}(w ; \boldsymbol{\lambda})\) can be computed, and JTransNet can then be trained using backpropagation and gradient descent to minimize this loss. Given a fixed \(\boldsymbol{\lambda}^t\), the gradient of \(C_\rho\) with respect to \(w\) is expressed as:
\begin{align*}
\label{pppooo}
\quad \nabla_{w} \mathcal{C}_{\rho}\left(w ; \boldsymbol{\lambda}^{t}\right)= &-\frac{1}{Y} \sum_{\ell=1}^{Y} \left[\sum_{i=1}^{m} \nabla_{w} p_{i \cdot }^{w}\left(\boldsymbol{v}^{(\ell)}\right)+ \sum_{j=1}^{n} \nabla_{w} p_{\cdot j}^{w}\left(\boldsymbol{v}^{(\ell)}\right)\right] \notag \\
& + \sum_{\ell=1}^{Y}\left[ \sum_{i=1}^{m}  \lambda_{i \cdot}^{t} g_{\ell, i \cdot}+\sum_{j=1}^{n} \lambda_{\cdot j}^{t} g_{\ell, \cdot j} \right] \notag \\
& +\rho \sum_{\ell=1}^{Y} \left[\sum_{i=1}^{m}  \widetilde{r g t}_{i \cdot}(w) g_{\ell, i \cdot }+ \sum_{j=1}^{n} \widetilde{r g t}_{\cdot j}(w) g_{\ell, \cdot j}\right]  \text{\hfill ,}
\end{align*}
where
\begin{equation*}
\left\{
\begin{split}
g_{\ell, i \cdot} &= \nabla_{w}\left[\max_{v_{i \cdot}^{\prime} \in V_{i \cdot}} u_{i \cdot}^{w}\left(v_{i \cdot}^{(\ell)} ;\left(v_{i \cdot}^{\prime}, \boldsymbol{v}_{-i \cdot}^{(\ell)}\right)\right) - u_{i \cdot}^{w}\left(v_{i \cdot}^{(\ell)} ; \boldsymbol{v}^{(\ell)}\right)\right] ,\\
g_{\ell, \cdot j} &= \nabla_{w}\left[\max_{v_{\cdot j}^{\prime} \in V_{\cdot j}} u_{\cdot j}^{w}\left(v_{\cdot j}^{(\ell)} ;\left(v_{\cdot j}^{\prime}, \boldsymbol{v}_{-\cdot j}^{(\ell)}\right)\right) - u_{\cdot j}^{w}\left(v_{\cdot j}^{(\ell)} ; \boldsymbol{v}^{(\ell)}\right)\right] .
\end{split}
\right.
\end{equation*}
The complete algorithm for training JTransNet is presented in Algorithm 1.

\begin{algorithm}[h!]
\caption{JTransNet Training} 
\begin{algorithmic}[1]
\STATE {\bf Input:} Minibatches 
$\mathcal{Y}_1, . . . , \mathcal{Y}_T$ of size Y \\
\STATE {\bf Parameters:} 
$\forall t \in \{1,\ldots,T\}, \rho_t > 0, \gamma > 0, \eta > 0, \Gamma \subseteq \mathbb{R}^+, T \in \mathbb{N},B \in \mathbb{N}
$ \\
\STATE {\bf Initialize:} 
$w^0 \in \mathbb{R}^{d}, \lambda^0 \in \mathbb{R}^{m+n}$
\FOR{$t=0$ to $T$} 
  \STATE Receive minibatch $\mathcal{Y}_t = \{v^{(1)}, \ldots, v^{(Y)}\}$ 
  \STATE Initialize misreport $v_{i \cdot}^{\prime(\ell)} \in V_{i \cdot},v_{\cdot j}^{\prime(\ell)} \in V_{\cdot j}, \forall \ell \in \{1,\ldots,Y\}, i \in M,  j \in N$
  \FOR{$r$ in $\Gamma$} 
    \STATE $\forall \ell \in \{1,\ldots,Y\}, i \in M,j \in N:$
    \STATE \noindent
\textbf{ $\, $ if} \(
\begin{aligned}
u_{i \cdot}^{w}\left(v_{i \cdot}^{(\ell)} ;\left({rv}_{i \cdot}^{(\ell)}, \boldsymbol{v}_{-i \cdot}^{(\ell)}\right)\right) &> 
u_{i \cdot}^{w}\left(v_{i \cdot}^{(\ell)} ;\left({v^{\prime}}_{i \cdot}^{(\ell)}, \boldsymbol{v}_{-i \cdot}^{(\ell)}\right)\right)
\end{aligned}
\)    
    \STATE $\, \,$ {\bf then} {$
            v_{i \cdot}^{\prime(\ell)}=r v_{i \cdot}^{\ell} $}
   \STATE $\, \,$ {\bf end if}
   \STATE  \noindent
\textbf{ $\, $ if} \(
\begin{aligned}
u_{\cdot j}^{w}\left(v_{\cdot j}^{(\ell)} ;\left({rv}_{\cdot j}^{(\ell)}, \boldsymbol{v}_{-\cdot j}^{(\ell)}\right)\right) &> 
u_{\cdot j}^{w}\left(v_{\cdot j}^{(\ell)} ;\left({v^{\prime}}_{\cdot j}^{(\ell)}, \boldsymbol{v}_{-\cdot j}^{(\ell)}\right)\right)
\end{aligned}
\)
   
    \STATE $\, \,$ {\bf then}  {$
            v_{\cdot j}^{\prime(\ell)}=r v_{\cdot j}^{\ell} $} 
    
   \STATE $\, \,$ {\bf end if}
  \ENDFOR
  \STATE Compute regret gradient:$\forall \ell \in [Y], i \in M, j \in N:$
  \STATE $\quad$ $g_{\ell, i \cdot}^{t}= \left. \nabla_{w}\left[u_{i \cdot}^{w}\left(v_{i \cdot}^{(\ell)} ;\left({v^{\prime}}_{i \cdot}^{(\ell)}, \boldsymbol{v}_{-i \cdot}^{(\ell)}\right)\right)-u_{i \cdot}^{w}\left(v_{i \cdot}^{(\ell)} ; \boldsymbol{v}^{(\ell)}\right)\right]\right|_{w=w^{t}}$
  \STATE $\quad$ $g_{\ell, \cdot j}^{t}= \left. \nabla_{w}\left[u_{\cdot j}^{w}\left(v_{\cdot j}^{(\ell)} ;\left({v^{\prime}}_{\cdot j}^{(\ell)}, \boldsymbol{v}_{-\cdot j}^{(\ell)}\right)\right)-u_{\cdot j}^{w}\left(v_{\cdot j}^{(\ell)} ; \boldsymbol{v}^{(\ell)}\right)\right]\right|_{w=w^{t}}$
  \STATE Compute Lagrangian gradient and update $w^t$:
  \STATE $\quad$ $w^{t+1} \leftarrow w^{t} - \eta \nabla_{w} \mathcal{C}_{\rho_{t}}\left(w^{t}, \lambda^{t}\right)$
  \STATE Update Lagrange multipliers once in $B$ iterations:
    \STATE $\quad$ {\bf if} {$t$ is a multiple of $B$} {\bf then} 
    \STATE $\quad$  $\quad$  $\lambda_{i \cdot}^{t+1} \leftarrow \lambda_{i \cdot}^{t}+\rho_{t} \widetilde{r g t}_{i \cdot}\left(w^{t+1}\right), \quad \forall i \in M$
     \STATE $\quad$  $\quad$  $\lambda_{\cdot j}^{t+1} \leftarrow \lambda_{\cdot j}^{t}+\rho_{t} \widetilde{r g t}_{\cdot j}\left(w^{t+1}\right), \quad \forall j \in N$
  \STATE $\quad$ {\bf else}
    \STATE $\quad$  $\quad$ $\boldsymbol{\lambda}^{t+1} \leftarrow \boldsymbol{\lambda}^t$
   \STATE $\quad$ {\bf end if}
\ENDFOR
\end{algorithmic}
\end{algorithm}

\section{Experiments}
\label{sec:exp}

In this section, we detail the empirical experiments conducted to demonstrate the effectiveness of JTransNet. 


\textbf{Baseline Methods.} The baseline methods used for comparison with JTransNet include: 

\begin{itemize} 

    \item \textbf{RegretNet} \cite{dutting2019optimal}, which can generate the optimal auction mechanisms satisfying near-DSIC and IR in traditional ad settings that only include stores.
    
    \item \textbf{VCG} \cite{vickrey1961counterspeculation, clarke1971multipart, groves1973incentives}, a widely used mechanism that satisfies DSIC and IR, which is applied to the joint ad setting in our experiments. It should be noted that our mechanism design objective differs from that of the VCG mechanism. We aim to design a mechanism that maximizes revenue, whereas the VCG mechanism is a mechanism that maximizes social welfare.
    
    \item \textbf{JAMA} \cite{ma2024joint}, a VCG-based  mechanism suitable for joint ad settings.

\end{itemize}








In our experiments, RegretNet is used for the traditional ad setting, while other methods are used for the joint ad setting. The difference between the joint and corresponding traditional ad settings is that the former includes both stores and brands as bidders, whereas the latter includes only stores. 



\textbf{Evaluation.} To comprehensively evaluate all the proposed methods, we adopt the following metrics:
\begin{itemize}
    \item The empirical revenue: 
    
    $rev := \frac{1}{L} \sum_{\ell=1}^{L} \left[\sum_{i=1}^{m} p_{i \cdot}^{w}\left(v^{(\ell)}\right) + \right.$ $\left. \sum_{j=1}^{n} p_{\cdot j}^{w}\left(v^{(\ell)}\right)\right]$.

    \item All the bidders' average empirical ex-post regret: 
    
    $\widehat{rgt} := \frac{1}{n+m} \left( \sum_{i=1}^{m} \widehat{rgt}_{i \cdot} + \sum_{j=1}^{n} \widehat{rgt}_{\cdot j} \right)$. 
    
    \item The empirical social welfare: 
    
    $sw := \frac{1}{L} \sum_{\ell=1}^{L} \sum_{c=1}^{C} \sum_{k=1}^{K} $ $\left( a_{i(c) \cdot k}^{(\ell)} \alpha_{k} v_{i(c) \cdot}^{(\ell)} + a_{\cdot j(c) k}^{(\ell)} \alpha_{k} v_{\cdot j(c)}^{(\ell)} \right)$. 
\end{itemize}

\textbf{Hyperparameters.} The training set sizes for the simulated data experiments and the real data experiments are $100000$ and approximately $230000$, respectively. The test set size for both the simulated data experiments and the real data experiments is $9984$. In addition, for the test of JTransNet, we use the coefficient set \(\Gamma = \{0, 0.05, 0.1, 0.15, \ldots, 1.45\}\) to obtain the optimal misreports for calculating the regret.

\subsection{Simulated Data Experiments}

\begin{table*}[]
\centering
\begin{tabular}{cllllll}
\hline
\multirow{2}{*}{\textbf{Method}} & {\multirow{2}{*}{{\begin{tabular}[c]{@{}c@{}}\multicolumn{1}{c}{\textbf{A: 2 $\times$ 4 $\times$ 1}}\\ \hspace{0.1cm} rev \quad \, sw \quad \,  rgt\end{tabular}}}} & {\multirow{2}{*}{{\begin{tabular}[c]{@{}c@{}}\multicolumn{1}{c}{\textbf{B: 3 $\times$ 3 $\times$ 2}}\\ \hspace{0.1cm} rev \quad \, sw \quad \, rgt\end{tabular}}}} & {\multirow{2}{*}{{\begin{tabular}[c]{@{}c@{}}\multicolumn{1}{c}{\textbf{C: 4 $\times$ 4 $\times$ 2}}\\ \hspace{0.1cm} rev \quad \, sw \quad \, rgt\end{tabular}}}} & {\multirow{2}{*}{{\begin{tabular}[c]{@{}c@{}}\multicolumn{1}{c}{\textbf{D: 4 $\times$ 4 $\times$ 3}}\\ \hspace{0.1cm} rev 
\quad \, sw \quad \, rgt\end{tabular}}}} \\
                                 & \multicolumn{1}{c}{}                                                                                      & \multicolumn{1}{c}{}                                                                                      & \multicolumn{1}{c}{}                                                                                      & \multicolumn{1}{c}{}                                                                                      & \multicolumn{1}{c}{}                                                                                      & \multicolumn{1}{c}{}                                             \\ \hline \\ [-10pt] \hline
RegretNet             &       0.294$^{\phantom{\dagger}}$  0.358 \, 0.004                                                                                                  &      0.403$^{\phantom{\dagger}}$   0.526        \, 0.003                                                                                             &                                      0.496$^{\phantom{\dagger}}$  0.571   \, 0.004                                                                 &     0.513$^{\phantom{\dagger}}$          0.590 \, 0.005                                                                                                      \\ \hline
VCG                        & 0.464$^{\phantom{\dagger}}$  \textbf{0.840}  \quad $-$                                                                                 & 0.306$^{\phantom{\dagger}}$    \textbf{1.008}  \quad $-$                                                                                 & 0.716$^{\phantom{\dagger}}$   \textbf{1.160}  \quad $-$                                                                                  & 0.725$^{\phantom{\dagger}}$   \textbf{1.228}   \quad $-$                                                                                                                                                                 \\ \hline 
JAMA                              &      0.501$^{\phantom{\dagger}}$  0.772  \quad \ $-$                                                                                                     &                  0.455$^{\phantom{\dagger}}$  0.884  \quad \ $-$                                                                                       &   0.717$^{\phantom{\dagger}}$   1.090  \quad \ $-$                                                                                                      &     0.728$^{\phantom{\dagger}}$   1.120  \quad \ $-$                                                                                                                                                                                                                                                                                                                                                   \\ \hline
JTransNet                          & \textbf{0.515$^{\dagger}$}\! \! \! 0.840 \  0.004                                                                                 & \textbf{0.491$^{\dagger}$}\! \! \! 1.008  \  0.003                                                                                 & \textbf{0.741$^{\dagger}$}\! \! \! 1.157 \ 0.004                                                                                 & \textbf{0.742$^{\dagger}$}\! \! \! 1.226  \  0.005                                                                                                                                                        \\ \hline
\end{tabular}
\caption{The experimental results for Settings A to D. `${\dagger}$` indicates that the revenue shows a statistically significant improvement compared to other methods in the paired t-test with a significance level of $p < 0.05$. }\label{tbl:1}
\end{table*}

\begin{table*}[]
\centering
\begin{tabular}{cllllll}
\hline
\multirow{2}{*}{\textbf{Method}} & {\multirow{2}{*}{{\begin{tabular}[c]{@{}c@{}}\multicolumn{1}{c}{\textbf{E1}}\\  rev \quad \quad sw \quad \ \ \ \, \ rgt\end{tabular}}}} & {\multirow{2}{*}{{\begin{tabular}[c]{@{}c@{}}\multicolumn{1}{c}{\textbf{E2}}\\  rev \quad \quad sw \quad \ \ \ \, \ rgt\end{tabular}}}} & {\multirow{2}{*}{{\begin{tabular}[c]{@{}c@{}}\multicolumn{1}{c}{\textbf{E3}} \\  rev 
\quad \quad sw \quad \ \ \ \, \  rgt\end{tabular}}}} \\
                                 & \multicolumn{1}{c}{}                                                                                      & \multicolumn{1}{c}{}                                                                                      & \multicolumn{1}{c}{}                                                                                      & \multicolumn{1}{c}{}                                                                                      & \multicolumn{1}{c}{}                                                                                      & \multicolumn{1}{c}{}                                             \\ \hline \\ [-10pt] \hline

VCG                             &  \!\!\!\!17.179 \,  \! \! \textbf{57.670}  \quad \, \  $-$                                                                                                                                                                     &  \!\!\!\!18.019 \,  \! \! \textbf{58.279}  \quad \, \ $-$                                                                                                      &  \!\!\!\!17.475  \,  \! \! \textbf{57.873}   \quad \, \ $-$                                                                                                                                                                                                            \\  \hline  

JAMA                             &  \!\!\!\!30.863  \,  \! \! 52.204  \quad  \, \, \! $-$                                                                                                                                                                     &  \!\!\!\!31.539  \,  \! \! 52.962  \quad \, \, \! $-$                                                                                                      &  \!\!\!\!30.985  \,  \! \! 52.464  \quad \, \, \! $-$

\\ \hline
JTransNet                          &\!\!\!\!\textbf{31.176$^{\dagger}$} \! \! 56.989 \,  $<$0.130                                                                                                                                                                & \!\!\!\!\textbf{32.048$^{\dagger}$} \! \! 57.549 \, $<$0.130                                                                                 & \!\!\!\!\textbf{31.537$^{\dagger}$} \! \! 57.131 \, $<$0.130                                                                                                                                        \\ \hline
\end{tabular}
\caption{The experimental results for real-world data. `${\dagger}$` indicates that the revenue shows a statistically significant improvement compared to other methods in the paired t-test with a significance level of $p < 0.05$. }\label{tbl:999}
\end{table*}

\begin{table*}[]
\centering
\begin{tabular}{cllllll}
\hline
\multirow{2}{*}{\textbf{Hyperparameter}} & {\multirow{2}{*}{{\begin{tabular}[c]{@{}c@{}}\multicolumn{1}{c}{\textbf{E1}}\\  rev \quad \quad sw \quad \ \ \ \, \ rgt\end{tabular}}}} & {\multirow{2}{*}{{\begin{tabular}[c]{@{}c@{}}\multicolumn{1}{c}{\textbf{E2}}\\  rev \quad \quad sw \quad \ \ \ \, \ rgt\end{tabular}}}} & {\multirow{2}{*}{{\begin{tabular}[c]{@{}c@{}}\multicolumn{1}{c}{\textbf{E3}} \\  rev 
\quad \quad sw \quad \ \ \ \, \  rgt\end{tabular}}}} \\
                                 & \multicolumn{1}{c}{}                                                                                      & \multicolumn{1}{c}{}                                                                                      & \multicolumn{1}{c}{}                                                                                      & \multicolumn{1}{c}{}                                                                                      & \multicolumn{1}{c}{}                                                                                      & \multicolumn{1}{c}{}                                             \\ \hline \\ [-10pt] \hline

$\tau = 0.0010$ $\rho = 0.60$                         &\!\!\!\! 31.176 \, \! \! 56.989 \,  $<$0.130                                                                                                                                                                & \!\!\!\! 32.048 \, \! \! 57.549 \, $<$0.130                                                                                 & \!\!\!\! 31.537 \, \! \! 57.131 \, $<$0.130                                                                                                                                                                                                         \\  \hline  

$\tau = 0.0010$ $\rho = 0.57$                          &\!\!\!\! 31.132 \, \! \! 56.884 \,  $<$0.130                                                                                                                                                                & \!\!\!\! 32.036 \, \! \! 57.615 \, $<$0.130                                                                                 & \!\!\!\! 31.408 \, \! \! 57.058 \, $<$0.130

\\ \hline
$\tau = 0.0009$ $\rho = 0.60$                          &\!\!\!\! 31.063 \, \! \! 57.083 \,  $<$0.130                                                                                                                                                                & \!\!\!\! 32.167 \, \! \! 57.516 \, $<$0.130                                                                                 & \!\!\!\! 31.187 \, \! \! 56.944 \, $<$0.130      \\ \hline

$\tau = 0.0011$ $\rho = 0.60$                          &\!\!\!\! 31.058 \, \! \! 56.651 \,  $<$0.130                                                                                                                                                                & \!\!\!\! 31.957 \, \! \! 57.436 \, $<$0.130                                                                                 & \!\!\!\! 31.462 \, \! \! 57.087 \, $<$0.130                                                                                                                                   \\ \hline
\end{tabular}
\caption{The real-world data experimental results of JTransNet with varying initial values of hyperparameter $\rho$ and varying values of hyperparameter $\tau$ (other hyperparameters fixed).}\label{tbl:666666}
\end{table*}

We conduct experiments under various settings to thoroughly evaluate the performance of JTransNet. The specific experimental settings are as follows: 

\begin{enumerate}
\item[(A)] $2$ stores and $4$ brands with $1$ ad position. The CTR is $\boldsymbol{\alpha}=(0.6)$. For all brands and stores, $v_{i \cdot} \sim U[0,1]$ and $v_{\cdot j} \sim U[0,1]$.

\item[(B)] $3$ stores and $3$ brands with $2$ ad positions. The CTRs are $\boldsymbol{\alpha}=(0.6, 0.2)$. For all brands and stores, $v_{i \cdot} \sim U[0,1]$ and $v_{\cdot j} \sim U[0,1]$.

\item[(C)] $4$ stores and $4$ brands with $2$ ad positions. The CTRs are $\boldsymbol{\alpha}=(0.6, 0.2)$. For all brands and stores, $v_{i \cdot} \sim U[0,1]$ and $v_{\cdot j} \sim U[0,1]$. 

\item[(D)] $4$ stores and $4$ brands with $3$ ad positions. The CTRs are $\boldsymbol{\alpha}=(0.6, 0.2,0.06)$. For all brands and stores, $v_{i \cdot} \sim U[0,1]$ and $v_{\cdot j} \sim U[0,1]$.
\end{enumerate}



\begin{figure}[h!]
\centering
\includegraphics[width=0.4\textwidth]{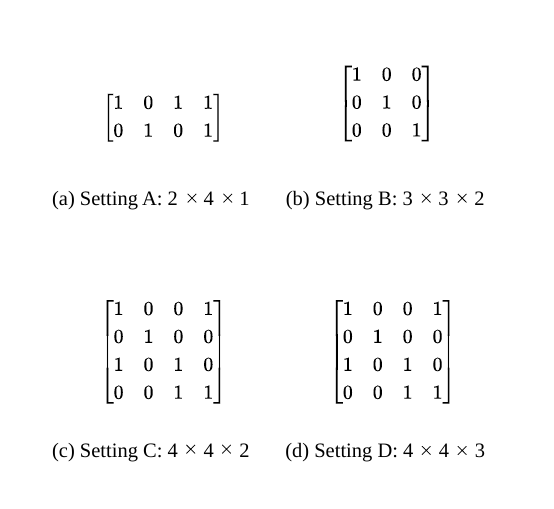}
\caption{The bundle relationships between store bidders and brand bidders for Settings $A$ to $D$. In the bundle relationship matrix, rows represent stores and columns represent brands.}
\label{fig:7}
\end{figure}

The bundle relationships of Settings $A$ to $D$ are shown separately in Figure \ref{fig:7}.

From Table \ref{tbl:1}, it can be seen that compared to all three baseline methods, especially JAMA which maximizes the revenue of VCG in joint auction scenarios, JTransNet achieves significantly higher revenue with very low regret. This demonstrates that JTransNet can generate anonymous and deterministic-allocation mechanisms that satisfy near-DSIC and IR while achieving high revenue. 
Furthermore, as shown in Table \ref{tbl:1}, JTransNet achieves a significant increase in revenue compared to VCG, which maximizes social welfare, while maintaining only a minimal loss in social welfare. This demonstrates JTransNet's excellent ability to balance revenue and social welfare.

\subsection{Real Data Experiments}

We utilize real-world auction log data from an e-commerce platform, which includes predicted CTRs, the relaxed joint relationships between brands and stores, as well as the bid information of each advertiser, to train and test JTransNet. We evaluate our model in a real-world setting with $10$ brands, $10$ stores, and $5$ positions. This setting reflects the current scenario on the e-commerce platform, where, after the stages of ad recall, coarse ranking and fine ranking, a maximum of $10$ bundles, comprising up to $10$ brands and $10$ stores, participate in the ad auction for the allocation of up to $5$ ad positions. 
We conduct three experiments on real-world data: E1, E2, and E3.

Since we have proven that the non-deterministic allocations generated by RegretNet contain many infeasible solutions, it cannot be applied in real advertising scenarios. Therefore, RegretNet is not included in the real data baselines. In Table \ref{tbl:999}, we present the performance of all methods except RegretNet on the test set across all real data experiments.
As shown in Table \ref{tbl:999}, in all real data experiments, JTransNet significantly outperforms all other baseline methods considered in terms of revenue while maintaining low regret.
This demonstrates the effectiveness of JTransNet in real-world scenarios. 
Furthermore, it is worth noting that even in real-world scenarios, JTransNet still achieves a substantial increase in revenue while incurring only a minor loss in social welfare compared with VCG.
This remarkable performance highlights JTransNet's superior ability to harmonize revenue and social welfare in the real data scenarios, making it an outstanding choice for real joint auction scenarios.

Additionally, in the real-data experiments, we evaluate the sensitivity of JTransNet to different initial values of the hyperparameter $\rho$ and to the hyperparameter $\tau$. 
As shown in Table \ref{tbl:666666}, for all real data experiments, while keeping the initial values of other hyperparameters unchanged, the revenue performance of JTransNet changes slightly after varying the initial value of $\rho$ and the value of $\tau$ within a certain range, demonstrating the robustness of JTransNet with respect to both parameters.





\subsection{Online A/B Test}



To further demonstrate the effectiveness of JTransNet, we deploy JTransNet in the joint auction scenario on a real e-commerce platform and conduct an A/B test to further validate the effectiveness of JTransNet. In real business deployment, a short inference time needs to be ensured to meet the platform’s online performance requirements. To achieve this, we deploy our JTransNet using a localized model service in the online mechanism server, which reduces the RPC (Remote Procedure Call) latency from cross-service access, keeping the average inference time of JTransNet under 10ms. This effectively supports the model’s launch. In our industrial deployment, we further measure the TP99 (99th percentile response time) of JTransNet to be 15 ms, meaning that 99\% of request response times are below 15 ms. Compared to the average inference time, TP99 better captures the tail latency of the model, reflecting the worst-case experience for a small fraction of requests.
In the online A/B test conducted over a period of one month with 40\% of the traffic, JTransNet achieved a substantial revenue increase of +1.87\% compared to the joint auction mechanism previously used by the e-commerce platform.
Currently, JTransNet has been fully deployed on the e-commerce platform and continues to generate high advertising revenue. All of these fully demonstrate the practical value and scalability of our model in large-scale industrial scenarios.

In practical industrial deployment, we have also considered whether to adopt an exact-IC architecture or a near-IC architecture. First, real-world scenarios differ from theoretical ones. Although we theoretically assume that advertisers always bid to maximize their utility, in real advertising scenarios, advertisers often cannot find the optimal misreporting strategy due to the cost of doing so, and instead use a relatively optimal bidding strategy as a substitute. In such practical scenarios, near-IC can almost guarantee truthful bidding by advertisers due to the cost of finding the optimal misreport. Moreover, in these cases, near-IC may achieve higher revenue than exact-IC by sacrificing a certain degree of incentive compatibility. Our experiments also demonstrate that JTransNet achieves significantly higher platform revenue compared to the exact-IC method JAMA.







\section{Conclusion}
\label{sec:conc}

In this paper, we propose an end-to-end neural network architecture, JTransNet, designed to generate optimal anonymous and deterministic-allocation joint auction mechanisms that satisfy DSIC and IR. Our JTransNet is the first AMD architecture for optimal near-DSIC auction design to be deployed on a real e-commerce platform, where online A/B testing demonstrates a substantial increase in revenue. Additionally, in both real and simulated data experiments, JTransNet achieves the significantly highest revenue among all compared methods. These results collectively demonstrate the effectiveness of JTransNet. 

In the future, it would be interesting to extend our deterministic allocation module to other AMD architectures, thereby promoting the widespread application of AMD architectures in the industry and further advancing the development of the advertising field. Another promising direction is to integrate contextual information or user preferences into deterministic-allocation and anonymous joint auction design.







\section*{Acknowledgments} 

Qi Qi is the corresponding author. This work was supported by National Natural Science Foundation of China (NO.62472428), Public Computing Cloud, Renmin University of China, the fund for building world-class universities (disciplines) of Renmin University of China, Meituan Inc. Fund, and the Qiushi Academic Project of Renmin University of China (NO.RUC25QSDL122).

\newpage
\clearpage
\bibliographystyle{unsrt}
\bibliography{main}

\clearpage
\appendix

\section{Proof of Theorem 1}
\label{ap1}

\subsection{An Example}

First, consider the scenario where 3 bundles compete for 2 ad slots under a joint scenario.  
Let the probability matrix \( S' \) be the output of the non-deterministic AMD method for each bundle and each slot.  
Each element \( s_{ck}' \) in \( S' \) represents the probability that the \( c \)-th bundle wins the \( k \)-th ad slot, where \( c \in \{1, 2, 3\} \) and \( k \in \{1, 2\} \).

According to practical requirements, the following two feasibility constraints must be satisfied:  
(a) a bundle can occupy no more than one slot, expressed as \( \sum_{k=1}^{K} s_{c k}' \leq 1 \) for all \( c \in \mathbb{C} \); and (b) a single slot can be assigned to no more than one bundle, expressed as \( \sum_{c=1}^{C} s_{c k}' \leq 1 \) for all \( k \in \mathbb{K} \).

Currently, the mainstream non-deterministic allocation AMD architecture satisfies feasibility constraints (a) and (b) by constructing a probability allocation matrix \( S' \), where both the row and column sums are less than or equal to 1. For any mainstream non-deterministic allocation AMD method, the set of allocation matrices it generates is exactly the set of all $S'$ matrices whose row and column sums are less than or equal to 1.
Thus, we have:
\[
s_{11}' + s_{12}' \leq 1, \quad s_{21}' + s_{22}' \leq 1, \quad s_{31}' + s_{32}' \leq 1 ,
\]
\[
s_{11}' + s_{21}' + s_{31}' \leq 1, \quad s_{12}' + s_{22}' + s_{32}' \leq 1 .
\]

For all results where all slots are allocated and feasibility constraints (a) and (b) are satisfied, there are 6 possible scenarios in total:

1. The 1st bundle is assigned to the 1st slot, and the 2nd bundle is assigned to the 2nd slot.  

2. The 1st bundle is assigned to the 1st slot, and the 3rd bundle is assigned to the 2nd slot.  

3. The 1st bundle is assigned to the 2nd slot, and the 2nd bundle is assigned to the 1st slot.  

4. The 1st bundle is assigned to the 2nd slot, and the 3rd bundle is assigned to the 1st slot.  

5. The 2nd bundle is assigned to the 1st slot, and the 3rd bundle is assigned to the 2nd slot.  

6. The 2nd bundle is assigned to the 2nd slot, and the 3rd bundle is assigned to the 1st slot.

Let \( Pr_1, Pr_2, \dots, Pr_6 \) represent the probabilities of scenarios 1, 2, ..., 6 occurring.  
To ensure that all probabilities in the allocation probability matrix \( S' \) are satisfied, the following 6 equations must hold:

\begin{align}
Pr_1 + Pr_2 &= s_{11}' ,\\
Pr_3 + Pr_4 &= s_{12}' ,\\
Pr_3 + Pr_5 &= s_{21}' ,\\
Pr_1 + Pr_6 &= s_{22}' ,\\
Pr_4 + Pr_6 &= s_{31}' ,\\
Pr_5 + Pr_2 &= s_{32}' .
\end{align}

Since there are only 6 possible scenarios, we have:  
\begin{align}
Pr_1 + Pr_2 + \cdots + Pr_6 &= 1 .
\end{align}

Substituting equations (5), (7), and (9) into (11) gives:
\[
s_{11}' + s_{21}' + s_{31}' = 1 ,
\]
and substituting equations (6), (8), and (10) into (11) gives:
\[
s_{12}' + s_{22}' + s_{32}' = 1 .
\]

Since \( s_{11}' + s_{21}' + s_{31}' \leq 1 \) and \( s_{12}' + s_{22}' + s_{32}' \leq 1 \),  
if constraints (a) and (b) are to be satisfied and all slots are allocated, many infeasible probability allocation matrices $S'$ will exist.  

\subsection{Proof in General Scenarios}

Now consider the more general case where \( C \) bundles compete for \( K \) ad slots in a joint scenario.  
We still use the matrix \( S' \) to represent the probability allocation matrix output by the non-deterministic AMD method. Each element \( s_{ck}' \) in \( S' \) represents the probability that the \( c \)-th bundle wins the \( k \)-th ad slot, where \( c \in \{1, \ldots, C\} \) and \( k \in \{1, \ldots, K\} \).



Since the mainstream non-deterministic allocation AMD architecture satisfies feasibility constraints (a) and (b) by constructing a probability allocation matrix \( S' \) whose row and column sums are less than or equal to 1, we have:


\begin{align}
\sum_{k=1}^{K} s_{c k}' \leq 1, \quad \forall c \in \{1, \ldots, C\},
\end{align}
and 
\begin{align}
\sum_{c=1}^{C} s_{c k}' \leq 1, \quad \forall k \in \{1, \ldots, K\}.
\end{align}

Let \( X \) be the number of deterministic allocation results that satisfy constraints (a) and (b) and where all ad slots are allocated.  
Let \(Pr_x \) represent the probability of the \( x \)-th outcome occurring, \( x \in \{1, 2, \dots, X\} \).  
Clearly, \( \sum_{x=1}^{X} Pr_x = 1 \).

Let \( s_{ck}' \) represent the probability that the \( c \)-th bundle wins the \( k \)-th ad slot.  
Let \( Pr^{ck} \) represent the probability set of a group of deterministic allocation results that realize probability \( s_{ck}' \).  
Let \( Pr \) represent the probability set of all possible deterministic allocation results.

For deterministic allocation results, the first slot can only be assigned to one of the bundles from 1 to \( C \),  
so the probability sets from \( Pr^{11} \) to \( Pr^{C1} \) are mutually exclusive.  
Since the first slot must be allocated, we have:
\[
Pr^{11} \cup Pr^{21} \cup \dots \cup Pr^{C1} = Pr ,
\]
so the sum of all the probabilities in the sets from \( Pr^{11} \) to \( Pr^{C1} \) equals 1:
\[
\sum_{c=1}^{C} s_{c1}' = 1 .
\]
Similarly,  
\[
\sum_{c=1}^{C} s_{c k}' = 1, \quad \forall k \in \{2, 3, \dots, K\}.
\]

According to formula (13), if feasibility constraints (a) and (b) are to be satisfied and all slots are allocated, many infeasible probability allocation matrices $S'$ will exist.  
If the allocation probabilities in \( S' \) are altered, it will affect DSIC and other economic properties.  
Moreover, as previously mentioned, the allocation and payment rules are generated by the AMD method, with the payment rules being contingent on the allocation rules. Therefore, modifying the allocation rules and the allocation probability matrix \( S' \) will fundamentally change the mechanism produced by the AMD method, leading to its failure.

Similarly, in more general traditional online advertising scenarios, there are also constraints similar to (a) and (b) that need to be satisfied: (a') each advertiser can be assigned at most one ad slot; (b') each ad slot can be assigned to at most one advertiser. Therefore, the mainstream non-deterministic allocation AMD architecture will still encounter the same problem.

\end{document}